\documentclass[fleqn,twoside]{article}

\usepackage{espcrc2}
\usepackage{graphicx}

\readRCS
$Id: espcrc2.tex,v 1.2 2004/02/24 11:22:11 spepping Exp $
\usepackage{graphicx}
\usepackage[figuresright]{rotating}

\newcommand{\AmS}{{\protect\the\textfont2
  A\kern-.1667em\lower.5ex\hbox{M}\kern-.125emS}}

\def\XSummer{{\sffamily\scshape XSummer}}
\def\Summer{{\sffamily\scshape Summer}}
\def\Form{{\sffamily\scshape Form}}
\def\Mathematica{{\sffamily\scshape Mathematica}}

\def\sign(#1){(\!-\!1)^{#1}}
\def\binom(#1,#2){ (\!\!
         \begin{array}{c} #1 \\ #2 \end{array}\!\! ) }
\def\z#1{{\zeta_{#1}}}

\title{
\vspace*{-35mm}
\rightline{
{\normalsize{DESY 05-185}}}
\vspace*{-2mm}
\rightline{
{\normalsize{SFB/CPP-05-60}}}
\vspace*{-2mm}
\rightline{\normalsize{August 2005}}
\vspace*{+20mm}
Symbolic Summation and Higher Orders in Perturbation Theory{\thanks{Presented at {\it{X International Workshop on Advanced Computing and Analysis Techniques in Physics Research}},
22th - 27th May 2005, Zeuthen (Germany).}}}

\author{S. Moch\address{Deutsches Elektronensynchrotron DESY \\
Platanenallee 6, D--15735 Zeuthen, Germany}}
       
\begin{document}
\pagestyle{empty}

\begin{abstract}
Higher orders in perturbation theory require the
calculation of Feynman integrals at multiple loops.
We report on an approach to systematically solve
Feynman integrals by means of symbolic summation 
and discuss the underlying algorithms.
Examples such as the non-planar vertex at two loops,
or integrals from the recent calculation of 
the three-loop QCD corrections to structure functions 
in deep-inelastic scattering are given.
\vspace{1pc}
\end{abstract}

\maketitle
\thispagestyle{empty}

\section{INTRODUCTION}

Symbolic summation amounts to finding a closed-form expression for 
a given sum or series.
Systematic studies have been pioneered by Euler~\cite{Euler:1775xx}, and
for specific sums, exact formulae have been known for a long time.
Today, general classes of sums, for example harmonic sums, 
have been investigated (see e.g. Refs.~\cite{Hoffman:refs})
and symbolic summation has further advanced through the development 
of algorithms suitable for computer algebra systems.
Here, the possibility to obtain exact solutions 
by means of recursive methods has lead to significant progress, 
for instance in the summation of rational or hypergeometric series, 
see e.g. Ref.~\cite{Petkovsek:1997ab}.

In quantum field theory, higher-order corrections in perturbation theory 
require the evaluation of Feynman diagrams, which describe 
real and virtual particles in a given scattering process.
In mathematical terms, Feynman diagrams are given as integrals over 
the loop momenta of the associated particle propagators. 
These integrals may depend on multiple scales and are usually divergent, 
thus requiring some regularization.
The standard choice is dimensional regularization, i.e. 
an analytical continuation of the dimensions of space-time from 4 to $D$, 
which keeps underlying gauge symmetries manifest.
Analytical expressions for Feynman integrals in $D$ dimensions may lead to transcendental 
or generalized hypergeometric functions, which have a series representation through nested sums 
with symbolic arguments. 
The main computational task is then to obtain the Laurent series upon expansion of the relevant 
functions in the small parameter $\epsilon=(D-4)/2$.

\section{ALGORITHMS}

The basic recursive definition of nested sums is given by~\cite{Moch:2001zr}
\begin{eqnarray}
\label{eq:def-S}
{\lefteqn{
S(n;m_1,...,m_k;x_1,...,x_k) = }}
\nonumber
\\& &
        \sum\limits_{j=1}^n {x_1^j \over j^{m_1}} S(j;m_2,...,m_k;x_2,...,x_k) \, .
\end{eqnarray}
where generally all $|x_i| \le 1$. The sum of all $m_i$ is called the weight of
the sum, while the index $k$ denotes the depth. 
This definition actually includes as special cases the 
classical polylogarithms, 
Nielsen functions, multiple and harmonic polylogarithms~\cite{Goncharov,%
Borwein,Remiddi:1999ew} in their 
series representations.
For all $x_i = 1$, the above definition reduces to harmonic
sums~\cite{Euler:1775xx,%
Zagier,Borwein:1996yq,Vermaseren:1998uu} 
and, if additionally the upper summation boundary $n \to \infty$, 
one recovers the (multiple) zeta values 
associated to Riemann's zeta-function~\cite{Hoffman:refs}.

As an important property the $S$-sums in Eq.~(\ref{eq:def-S}) 
obey the well-known algebra of multiplication.
Specifically, any product
\begin{eqnarray}
\label{eq:prod-S}
{\lefteqn{
  S(n;m_1,...,m_k;x_1,...,x_k)}}
\nonumber
\\& &
   \times S(n;m^{\prime}_1,...,m^{\prime}_l;x^{\prime}_1,...,x^{\prime}_l)
\, ,
\end{eqnarray}
can be expressed again as a sum of single nested sums, 
hence in a canonical form, 
which is an important feature for practical applications.
The underlying algebraic structure in Eq.~(\ref{eq:prod-S}) is a Hopf algebra,
being realized as a quasi-shuffle algebra here, 
see e.g. Refs.~\cite{Moch:2001zr,Hoffman,Weinzierl:2003ub,Blumlein:2003gb} .
The algorithm  can be implemented very efficiently on a computer, 
see e.g. Refs.~\cite{Vermaseren:1998uu,Moch:2005uc}.

For the manipulation of the $S$-sums, we classify certain 
types of transcendental sums. 
All sums in these classes 
can be solved recursively, i.e. they can be expressed in canonical form.
The underlying algorithms realize a creative telescoping. 
They either reduce successively the depth or the weight of the inner sum, 
so that eventually the inner nestings vanish. 
Finally, the results can be written in the basis of Eq.~(\ref{eq:def-S}), 
(as $S$-sums with upper summation limit $n$) or as 
multiple polylogarithms (which are $S$-sums to infinity).

Besides the quasi-shuffle algebra of multiplication in Eq.~(\ref{eq:prod-S}),
the procedure relies on algebraic manipulations, 
such as partial fractioning of denominators, 
shifts of the summation ranges and synchronization of summation boundaries 
of the individual sums.

Specifically, we consider convolutions, 
\begin{eqnarray}
\label{eq:conv}
{\lefteqn{
  \sum\limits_{j=1}^{n-1} 
  {x_1^j \over j^{m_1}} S(j;m_2,...,m_k;x_2,...,x_k) }}
\\& &
   \times {(x^{\prime}_1)^{n-j} \over (n-j)^{m^{\prime}_1}} 
    S(n-j;m^{\prime}_2,...,m^{\prime}_l;x^{\prime}_2,...,x^{\prime}_l)
\, ,
\nonumber
\end{eqnarray}
conjugations, 
\begin{eqnarray}
\label{eq:conj}
{\lefteqn{
  - \sum\limits_{j=1}^{n}\, 
  \left( \begin{array}{c} n \\ j \end{array} \right)\, 
  (-1)^j}}
\nonumber
\\& &
  \times {x_1^j \over j^{m_1}} S(j;m_2,...,m_k;x_2,...,x_k)\, ,\,\,
\end{eqnarray}
and binomial convolutions, 
\begin{eqnarray}
\label{eq:binoconv}
{\lefteqn{
  - \sum\limits_{j=1}^{n-1}\, 
  \left(\begin{array}{c} n \\ j \end{array} \right)\, 
  (-1)^j}}
\nonumber
\\& &
  \times {x_1^j \over j^{m_1}} S(j;m_2,...,m_k;x_2,...,x_k)
\\& &
  \times {(x^\prime_1)^{n-j} \over (n-j)^{m^{\prime}_1}} S(n-j;m^{\prime}_2,...,m^{\prime}_l;x^{\prime}_2,...,x^{\prime}_l)
\, .\,\,
\nonumber
\end{eqnarray}
In all cases, the upper summation boundary should 
be consistent with the defining range of the binomials and the $S$-sums.

\section{APPLICATIONS}

In perturbation theory, one way to classify Feynman integrals is 
according to the number of scales, i.e. the number of non-vanishing 
scalar products of external momenta or particle masses.
According to this criterion, analytical expressions in $D$ dimensions 
for the Laurent series in $\epsilon$ either lead to 
transcendental numbers like (multiple) zeta values or 
(multiple) polylogarithms.
Other classification criteria are, of course, 
the topology of a Feynman integral (number of loops and external legs).

\subsection{One-scale problems}

\begin{figure}[htb]
\begin{center}
\includegraphics[width=4.0cm]{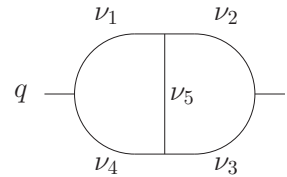}
\end{center}
\vspace*{-12mm}
\caption{The two-loop self-energy ${\rm T1}$.}
\label{fig:1}
\end{figure}

Prominent examples of one-scale problems are 
massless two-point functions~\cite{Broadhurst:2002gb,%
Bierenbaum:2003ud,Bekavac:2005xs}.
In particular, the massless two-loop self-energy ${\rm T1}$
has not only been of practical importance from a phenomenological perspective,
but received also quite some interest from number theorists.
For arbitrary powers of propgators, it is given by 
\begin{eqnarray}
  \label{eq:topo-T1}
  {\lefteqn{
      {\rm T1}(\nu_1,\nu_2,\nu_3,\nu_4,\nu_5) \,=\, 
      \int \frac{{\hbox{d}}^D p_1}{(2\pi)^D} \,
      \int \frac{{\hbox{d}}^D p_2}{(2\pi)^D} 
    }}
  \nonumber \\
  & & \times\, 
  {1 \over 
    \left(p_1^2\right)^{\nu_1}
    \left(p_2^2\right)^{\nu_2}
    \left(p_3^2\right)^{\nu_3}
    \left(p_4^2\right)^{\nu_4}
    \left(p_5^2\right)^{\nu_5}
  },
\end{eqnarray}
where $p_3=p_2-q$, $p_4=p_1-q$, $p_5=p_1-p_2$.
Graphically, it is displayed in Fig.~\ref{fig:1}.
\vspace*{-8mm}
\vspace*{-8mm}

Here the interesting question has been which types of (transcendental) 
numbers appear in the $\epsilon$-expansion of this integral.
For powers of the propagators of the form $\nu_i=1+\epsilon$, 
$i = 1,\dots,5$ it was known
from explicit calculations up to the $\epsilon^9$-term 
(heavily relying on symmetry properties) 
that multiple zeta values occur~\cite{Broadhurst:2002gb}.
However, it was unclear whether this suffices to all orders in $\epsilon$. 
Eventually, by deriving a double sum of (generalized) hypergeometric type,  
it was proven that multiple zeta values are indeed 
sufficient~\cite{Bierenbaum:2003ud}. 

Currently, with the help of symbolic summation 
the $\epsilon$-expansion is known to the 
$\epsilon^{13}$-term~\cite{Vermaseren:MZVtabs}.
The depth is limited by the fact 
that harmonic sums in infinity 
are expressed in a basis of transcendental numbers 
only up to weight 16.

\begin{figure}[htb]
\begin{center}
\includegraphics[width=5.0cm]{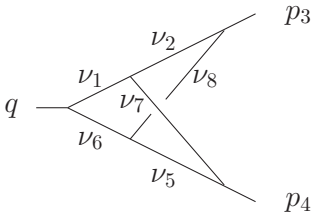}
\end{center}
\vspace*{-12mm}
\caption{The non-planar two-loop vertex $V_{\rm NO}$.}
\label{fig:2}
\end{figure}

Another example for a one-scale problem, that received attention 
recently~\cite{Moch:2005id,Gehrmann:2005pd} 
is the non-planar vertex at two loops, 
which enters in calculations of the quark and gluon form 
factors~\cite{Moch:2005id,Moch:2005tm} in QCD.
Here the basic integral (displayed in Fig.~\ref{fig:2})
is given by 
\begin{eqnarray}
  \label{eq:novertex}
  {\lefteqn{
      V_{\rm NO}(\nu_1,\nu_2,\nu_5,\nu_6,\nu_7,\nu_8) \,=\, 
      \int \frac{{\hbox{d}}^D p_1}{(2\pi)^D} \,
      \int \frac{{\hbox{d}}^D p_2}{(2\pi)^D} 
    }}
  \nonumber \\
  & & \times\, 
  {1 \over 
    \left(p_1^2\right)^{\nu_1}
    \left(p_2^2\right)^{\nu_2}
    \left(p_5^2\right)^{\nu_5}
    \left(p_6^2\right)^{\nu_6}
    \left(p_7^2\right)^{\nu_7}
    \left(p_8^2\right)^{\nu_8}
  },
\end{eqnarray}
where $p_5=p_4-p_7$, $p_6=p_1-q$, $p_7=p_2-p_1$, 
$p_8=p_2-p_3$ and $q^2 = (p_3-p_4)^2$.
\vspace*{-8mm}
\vspace*{-8mm}

For general powers of propagators, $V_{\rm NO}$ 
can be written as a double sum over Gamma functions,
and if all $\nu_i=1$, an expression 
in terms of hypergeometric functions 
$_3F_2$ and $_4F_3$ has been given in Ref.~\cite{Gehrmann:2005pd}.
After expansion, the sum can be solved in terms of the Riemann zeta 
function to any order in $\epsilon$ using the algorithms for harmonic sums
\cite{Vermaseren:1998uu,Moch:2005uc} 
coded, as all our symbolic manipulations, in \Form\ 
\cite{Vermaseren:2000nd}.
We find the following expansion to 
order $\epsilon^5$,
\begin{eqnarray}
  \label{eq:novertex-ep}
{\lefteqn{
V_{\rm NO}(1,1,1,1,1,1) \,=\,}}
\nonumber \\ & & 
S_\Gamma^2 \, \left(-q^2\right)^{-2-2\epsilon}\, \Bigg[
       - {1 \over \epsilon^4} 
       + {5 \* \z2  \over \epsilon^2} 
       + {27 \* \z3 \over \epsilon} 
\nonumber \\ & &
       + 23 \* \z2^2
       - \epsilon  \*  ( 48 \* \z2 \* \z3 - 117 \* \z5 )
       + \epsilon^2  \*  \bigl( {456 \over 35} \* \z2^3 
\nonumber \\ & &
                     - 267 \* \z3^2 \bigr)
       - \epsilon^3  \*  \bigl( {1962 \over 5} \* \z2^2 \* \z3 
                     + 240 \* \z2 \* \z5 + 6 \* \z7 \bigr)
\nonumber \\ & &
       - \epsilon^4  \*  \bigl( {3219 \over 7} \* \z2^4 - 264 \* \z2 \* \z3^2 
                   + 2466 \* \z3 \* \z5 
\nonumber \\ & &
                   - 264 \* \z{5,3} \bigr)
       - \epsilon^5  \*  \bigl( {2832 \over 5} \* \z2^3 \* \z3 
                     + 2718 \* \z2^2 \* \z5 
\nonumber \\ & &
                     + 1218 \* \z2 \* \z7 + 1626 \* \z3^3 
                     + {20777 \over 3} \* \z9 \bigr) \Bigg] \, ,
\end{eqnarray}
where we have taken out the usual ${\overline{\mbox{MS}}}$-scheme 
factor 
\begin{equation}
  \label{eq:pre}
S_\Gamma = {(4\pi)^{-2+\epsilon} \over \Gamma(1-\epsilon)}\, .
\end{equation}

This completes the section of examples with one-scale.

\subsection{Two-scale problems}

Nice examples of two-scale problems are provided by the 
recent calculation of the three loop corrections 
in Quantum Chromodynamics (QCD) 
to the structure functions of deep-inelastic scattering~\cite{Moch:2004pa,%
Vogt:2004mw,Vermaseren:2005qc}.
Here, the two scales are the virtuality of the exchanged 
gauge boson $Q^2 = -q^2$ and the scalar product of 
the boson's and nucleon's momenta, $2 p \cdot q$,
both combining to Bjorken's dimensionless 
variable $x = Q^2/(2p \cdot q)$.

The Feynman integrals under consideration can be expressed in 
nested sums and solved with the help of symbolic summation as 
follows.
Imagine a mapping of a given integral $I(x)$ depending on $x$, 
$0 \le x \le 1$ to the space of discrete variables $I(N)$, $N \in
{\mathbf N}$,
which is accomplished by means of an integral transformation, 
e.g. a Mellin transformation.
Then one can obtain difference equations for the Feynman integral $I(N)$, 
which may be written as~\cite{Moch:1999eb}
\begin{eqnarray}
\label{eq:diffeq}
{\lefteqn{
  a_0(N)\, I(N) + a_1(N)\,  I(N-1) + \ldots 
}}
\nonumber
\\& &
+ a_m(N)\,  I(N-m) \: = \: G(N)\, ,
\end{eqnarray}
where $G(N)$ is some inhomogeneous term and $a_i$ are some coefficients 
depending on $N$ (and perhaps on $\epsilon$). 
The solution of Eq.~(\ref{eq:diffeq}) needs $m$ 
boundary conditions $I(0), \dots, I(m-1)$.

Single-step difference equations can be summed up analytically in closed form.
Suppose we have the equation
\begin{eqnarray}
\label{eq:singlestepdiffeq}
        a_0(N)\, I(N) - a_1(N)\, I(N-1) \: = \: G(N)\, ,
\end{eqnarray}
then its solution will be
\begin{eqnarray}
\label{eq:firstsol}
{\lefteqn{
        I(N) = \frac{\prod_{j=1}^N a_1(j)}{\prod_{j=1}^N a_0(j)}\, I(0)}}
\nonumber
\\& &
                + \sum_{i=1}^N\frac{\prod_{j=i+1}^N a_1(j)}{\prod_{j=i}^N
                a_0(j)}\, G(i)\, .
\end{eqnarray}
In the case that the functions $a_i$ can be factorized in linear
polynomials of the type $N + m + n\, \epsilon$ with $m,n$ being integer and 
$N$ being symbolic, the products can be written as combinations of Gamma functions. 
In the presence of parametric dependence on $\epsilon$ the Gamma functions
should be expanded around $\epsilon = 0$, 
leading to factorials and harmonic sums.
If the function $G(N)$ is expressed as a Laurent series in $\epsilon$ with the 
coefficients being combinations of harmonic sums in $N+m$ and powers of 
$N+m$, $m$ being a fixed integer, the sum in Eq.~(\ref{eq:firstsol}) 
can be done and $I(N)$ will be a combination of harmonic sums in $N+k$ and 
powers of $N+k$ with $k$ being a fixed integer.

Eq.~(\ref{eq:diffeq}) is an example of a recursion for Feynman integrals 
with dependence on symbolic parameters.
Solutions such as Eq.~(\ref{eq:firstsol}) allow for an efficient
implementation in computer algebra systems like \Form\ 
\cite{Vermaseren:2000nd}  
resulting in a largely automatic build-up of nested sums.
For calculation of QCD corrections to structure functions 
mentioned above, a systematic evaluation of nested sums was 
required for all integrals occurring 
in approximately 10.000 Feynman diagrams. 
Because of the expressions being of excessive size at intermediate stages 
this task was well suited for the computer algebra system \Form\ 
and the \Summer\ package~\cite{Vermaseren:1998uu} for nested sums.

\subsection{Multi-scale problems}

Multi-scale problems arise in the calculation of 
cross sections with more kinematical invariants, like e.g. 
jet cross sections.
The methods and algorithms for generalized sums 
have already been used in full-fledged 
QCD calculations, for instance in 
the evaluation of higher order corrections to 
$e^+ e^- \to 3 \mbox{jets}$ \cite{Moch:2002hm}.

In general, Eqs.~(\ref{eq:conv})--(\ref{eq:binoconv}) may 
also be used to expand higher transcendental functions 
in a small parameter around integer values. 
Starting from the series representation of, e.g. 
the first Appell function
\begin{eqnarray}
\label{eq:appell1}
{\lefteqn{
F_1(a,b_1,b_2;c;x_1,x_2) \,=\,}}
\nonumber\\
& &
 \sum\limits_{m_1=0}^\infty \sum\limits_{m_2=0}^\infty
 \frac{a^{\overline{m_1+m_2}} b_1^{\overline{m_1}} b_2^{\overline{m_2}}}{c^{\overline{m_1+m_2}}} 
 \frac{x_1^{m_1}}{m_1!}
 \frac{x_2^{m_2}}{m_2!} \, ,
\end{eqnarray}
or the second Appell function
\begin{eqnarray}
\label{eq:appell2}
{\lefteqn{
F_2(a,b_1,b_2;c_1,c_2;x_1,x_2) \,=\,}}
\nonumber\\
& &
\sum\limits_{m_1=0}^\infty \sum\limits_{m_2=0}^\infty
 \frac{a^{\overline{m_1+m_2}} b_1^{\overline{m_1}} b_2^{\overline{m_2}}}{c_1^{\overline{m_1}} c_2^{\overline{m_2}}} 
 \frac{x_1^{m_1}}{m_1!}
 \frac{x_2^{m_2}}{m_2!} \, ,
\end{eqnarray}
we see that Eqs.~(\ref{eq:conv})--(\ref{eq:binoconv}) apply if
the expansion parameter $\epsilon$ occurs in the argument 
of the rising factorials (Pochhammer symbols), defined as
\begin{eqnarray}
f^{\overline{m}} &=& f(x)f(x+1) \dots f(x+m-1) \, .
\end{eqnarray}

It should also be stressed at this point, that although the 
definition of the $S$-sums in Eq.~(\ref{eq:def-S}) is very 
general, the specific algorithms for convolution, conjugation 
etc. are subject 
more restrictive assumptions.

In particular, the algorithms underlying the evaluation 
of Eqs.~(\ref{eq:conv})--(\ref{eq:binoconv}) 
do rely on the fact that the modulus of the summation index 
in the argument of the $S$-sums or in the denominators 
is always one.
This changes, if for instance hypergeometric functions $_JF_{J-1}$ 
(or more generally Gamma-functions) are expanded around 
half-integer values.
Such a situation occurs for example in the calculation of massive 
higher loop integrals in Bhabha scattering~\cite{Czakon:2004wm}.

Some extensions of the summation algorithms for expansion around 
rational values, leading e.g. to binomial sums and 
inverse binomial sums are discussed in 
Ref.~\cite{Davydychev:2003mv,Weinzierl:2004bn}.

\section{CONCLUSION}

Symbolic summation has advanced to an important method 
for the calculation of higher order corrections in 
perturbative quantum field theory.
The field has seen significant progress during the past years 
and we have given various examples from complete 
calculations, e.g. the recent evaluation of the third-order contributions 
in perturbative QCD to the structure functions of deep-inelastic 
scattering~\cite{Moch:2004pa,Vogt:2004mw,Vermaseren:2005qc}.
These cutting edge calculations show that the method of 
symbolic summation provides very powerful means for the 
practical computations of Feynman diagrams.

In closing, we note that all practical applications do heavily rely on 
computer algebra implementations of the algorithms discussed here.
In the symbolic manipulation program \Form\ \cite{Vermaseren:2000nd}, 
which is a fast and efficient computer algebra system to handle 
large expressions,
harmonic sums can be manipulated with the \Summer\ 
package~\cite{Vermaseren:1998uu}.
For the $S$-sums of Eq.~(\ref{eq:def-S}) 
there exists an extension, the \XSummer\ package~\cite{Moch:2005uc} 
in \Form\ , 
which implements algorithms of 
Eqs.~(\ref{eq:conv})--(\ref{eq:binoconv}).
As an alternative within the GiNaC framework~\cite{Bauer:2000cp}
the package {\sc nestedsums} \cite{Weinzierl:2002hv} provides 
similar functionalities.
Very recently, also Ref.~\cite{Huber:2005yg} appeared, 
which limits itself to the problem of expanding 
hypergeometric functions $_JF_{J-1}$ around 
integer parameters to arbitrary order and provides 
an implementation in \Mathematica\ .

We believe all these packages may also be useful for a larger community.

\end{document}